# String Art: Circle Drawing Using Straight Lines

Sankar K and Sarad AV

AU-KBC Research Centre,
MIT Campus of Anna University, Chennai, India
`{sankar@au-kbc.org, avsarad@au-kbc.org}`

**Abstract.** An algorithm to generate the locus of a circle using the intersection points of straight lines is proposed. The pixels on the circle are plotted independent of one another and the operations involved in finding the locus of the circle from the intersection of straight lines are parallelizable. Integer only arithmetic and algorithmic optimizations are used for speedup. The proposed algorithm makes use of an envelope to form a parabolic arc which is consequent transformed into a circle. The use of parabolic arcs for the transformation results in higher pixel errors as the radius of the circle to be drawn increases. At its current state, the algorithm presented may be suitable only for generating circles for string art.

## 1  Introduction

Circle drawing algorithms finds numerous applications in computer aided designing, gaming and visualizations, astronomy and computer simulations. An algorithm that generates a circle with all its tangents is considered as a bonus. A number of algorithms, in particular Bresenham's circle algorithm [1] and [2-5] are popular in circle drawing [6-7]. Parallelization of circle drawing algorithms find applications in high end graphics systems where each processor is assigned a specific task locally and their interactions with one another is synchronized globally for faster display[8-9]. An algorithm for the digital display of a circle using the intersection of straight lines, the intersections forming the locus of the circle is proposed. The line intersections are independently computable and hence the arithmetic operations involved in computing the line intersections are parallelized in a Multiple Instruction stream, Multiple Data stream (MIMD)[10] environment, where the processors operate independently and use a shared memory. The key idea is to look at the intersection of straight lines under certain constraints in the first quadrant of a two dimensional Cartesian coordinate system. Many graphic display devices place the origin of coordinates at the top left corner of the display, with X coordinate lengths increasing horizontally to the right of the display and the Y coordinate length increasing vertically down the display. Hence, the origin of coordinates for the discussion henceforth is the top left corner of the display.

Section 2 discusses the idea behind the proposed circle drawing algorithm and Section 2.1 involves improving computational efficiency and removing floating point operations. Section 3 discusses the necessary transformations required to generate the locus of the circle and Section 3.1 gives the experimental results. Section 4 gives the pseudo code for the circle drawing algorithm. Section 4.1 and 4.2 discusses the error bound estimates and the analysis of the algorithm respectively. Section 5 concludes the main results of the paper.

## 2  Line Intersection

The circle drawing algorithm is a result of the parabolic curve from an envelope [17] obtained by adjacent intersecting lines in the following setup:

*Step One*: With respect to the origin, divide the X and Y coordinate axis into an equal number of units ($i= 0, 1, 2… n$). Next, using straight lines join the points ((0, $n$), (1, 0)), ((0, $n$-1) (2, 0)). . . ((0, 1), ($n$, 0)) to yield lines $k_1, k_2. . . k_n$ respectively. Find the set of line intersection points $K=\{ (k_1, x=0),(k_2,k_1),(k_3,k_2),… (k_n,k_{n-1}),(k_n,y=0) \}$ where x=0 and y=0 are respectively, the vertical and horizontal line passing through the origin.

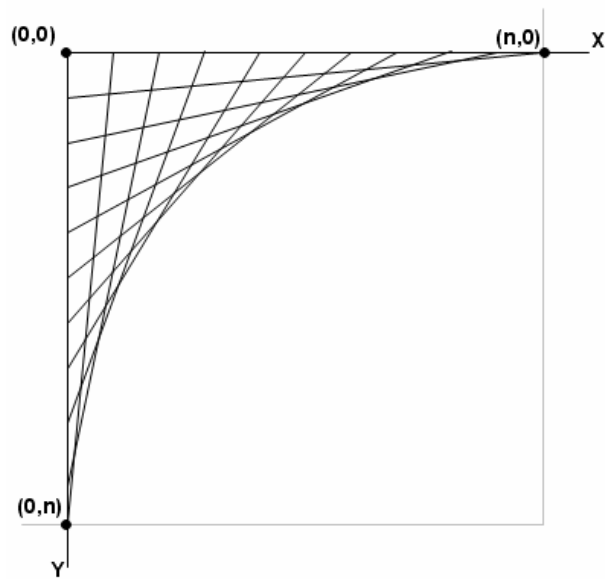

**Fig. 1.** Intersecting lines

The lines drawn in Fig. 1. are an illustration of the main idea. For the circle drawing algorithm, we are only concerned with the equations of these lines to find their intersections. The equation of the line joining points $(x_1, y_1)$ and $(x_2, y_2)$ is given by $y = m*x + c$ where $m$ is the slope of the line and $c$, its Y-intercept. The intersection point between adjacent lines is used to find the points on the locus of the circle. It is hence essential that the arithmetic operations are computationally light weight. *Step Two* discusses the simplifications involved.

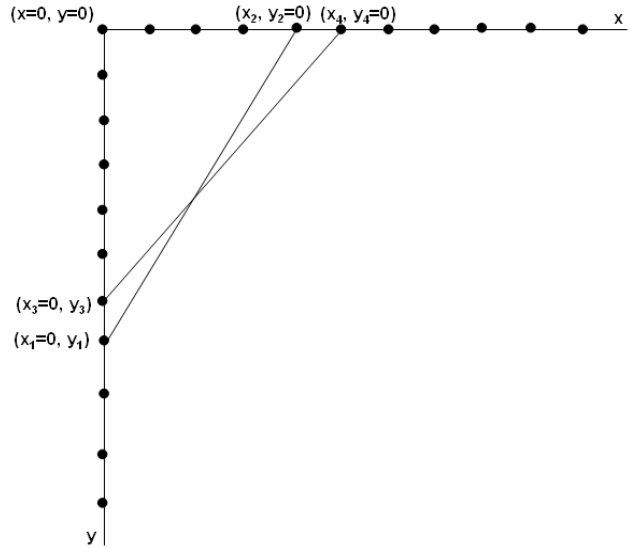

**Fig. 2.** Two adjacent lines

*Step Two*: Let the y-intercepts of the lines joining $((x_1, y_1), (x_2, y_2))$ and $((x_3, y_3), (x_4, y_4))$ in Fig. 2. be

$$c_1 = y_1 - m_1 * x_1. \tag{1}$$

and

$$c_2 = y_3 - m_2 * x_3. \tag{2}$$

Given two adjacent lines $-m_1*x + 1*y = c_1$ and $-m_2*x + 1*y = c_2$ from set $K$, solve this set of equations using Cramer's rule to find

$$x = (c_1 - c_2) / (m_2 - m_1). \tag{3}$$

Further,

$$(c_1 - c_2) = -1. \tag{4}$$

Find $m_2 - m_1 = [\,(y_3-y_4)/(x_3-x_4)\,] - [\,(y_1-y_2)/(x_1-x_2)\,]$.

$$m_2 - m_1 = [ (y_3-y_4)*(x_1-x_2) - (x_3-x_4)*(y_1-y_2) ] / [ (x_3-x_4)*(x_1-x_2) ]. \quad (5)$$

By equation 4 and equation 5, equation 3 transforms to

$$x = -[ (x_3-x_4)*(x_1-x_2) ] / [ (y_3-y_4)*(x_1-x_2) - (x_3-x_4)*(y_1-y_2) ]. \quad (6)$$

Since $x_1=x_3=0$ and $y_2=y_4=0$; on substitution in equation 6

$$x = [x_2*x_4] / [(x_4*y_1)-(x_2*y_3)]. \quad (7)$$

Since $x_4=x_2+1$ and $y_1=y_3+1$, equation 7 transforms to

$$x = [x_2*(x_2+1)] / [ x_4*(y_3+1) - (x_2*y_3) ]. \quad (8)$$

$$x = [x_2*(x_2+1)] / [ x_4*y_3 - x_2*y_3 + x_4 ].$$

$$x = [x_2*(x_2+1)] / [ y_3*(x_4 - x_2) + x_4 ]. \quad (9)$$

Since $x_4-x_2=1$; equation 9 transforms to

$$x = [x_2*(x_2+1)] / [ y_3 + x_4 ]. \quad (10)$$

The simplification of equation 3 to yield equation 10 eliminates the need to find the slopes and significantly reduces the time complexity of the arithmetic operations required to find $x$. Once $x$ is found by equation 10, find

$$y = m_2*x + c_2. \quad (11)$$

Substituting the value of $c_2$ from equation 2 in equation 11

$$y = m_2*x + y_3 - m_2*x_3. \quad (12)$$

Since $x_3=0$ and $y_4=0$, equation 12 transforms to

$$y = (-y_3 / x_4)*x + y_3. \quad (13)$$

$$y = y_3*[ 1 - (x/x_4) ]. \quad (14)$$

$$y = [ y_3* (x_4 - x) ] / x_4. \quad (15)$$

The simplification of equation 11 to yield equation 15 eliminates the need to find the slopes and significantly reduces the time complexity of the arithmetic operations required to find $y$.

## 2.1 Computational Speedup

Floating point arithmetic is computationally intensive and is eliminated by using the binary division algorithm. Multiplications are sped up by employing the modified Booth's algorithm and a Wallace tree. Let $k = \lfloor \log_2 n \rfloor + 1$ where $n$ is a base 10 integer. The time complexity to multiply two $k$ bit integers of the same size is $O(k^2)$ [11]. On parallel computer architecture, the time complexity of multiplication can be improved to O($k$) by using the modified Booth's algorithm with a Wallace tree [12-14]. The run of 1's and 0's in the bit string of the multiplier determines the speedup in multiplication for the modified Booth's algorithm and is variable from one multiplier to another. The parallelism due to the Wallace tree algorithm puts multiplication in the complexity class NC and the speed up in multiplication is close to the addition of two $k$ bit integers. The time complexity to divide two $k$ bit integers of the same size is O($k^2$) [11]. The only floating operation due to division is eliminated by using the binary division algorithm. It is essential to first compute the numerator and denominator separately and then carry out division to prevent loss of precision. The resulting quotient and remainder of binary division is stored separately in register/memory locations. Let << be the binary left shift operator. The following operations are carried out:

1. Set (remainder)$_2$ ← (remainder)$_2$<<1 bit
2. If (remainder)$_2$ ≥(divisor)$_2$
    set (quotient)$_2$←(quotient+1)$_2$
  else
    set (quotient)$_2$←(quotient)$_2$

Example. 31 = (1111)$_2$ and 10 = (1010)$_2$. Then 31/10 is computed as follows:

```
1010 | 11111 |  quotient=11
       1010
       -------
       01011 –
        1010
       ------------
        0001 = remainder
```

Since (1)$_2$<<1 = (10)$_2$ < (divisor)$_2$, (quotient)$_2$←(quotient)$_2$.

This is equivalent to finding the floor or ceiling function of (remainder/divisor) without the use of floating point arithmetic. If (remainder/divisor)<0.5, the floor function is computed, otherwise the ceiling function is computed. It is further possible to reduce the time spent on division by using a divide and conquer[15] approach on the binary division algorithm and puts division in O($k.\log k$). Once the equidistant points on the X and Y axis are determined, the operations required in finding the equation of the line and solving for the point of intersection of a pair of lines can be easily parallelized to $n$ parts for speed up.

## 3   Transformation and Experimental Setup

This section discusses the transformation and experimental setup required to collect the points that lie on the circle and its display on a digital display device.

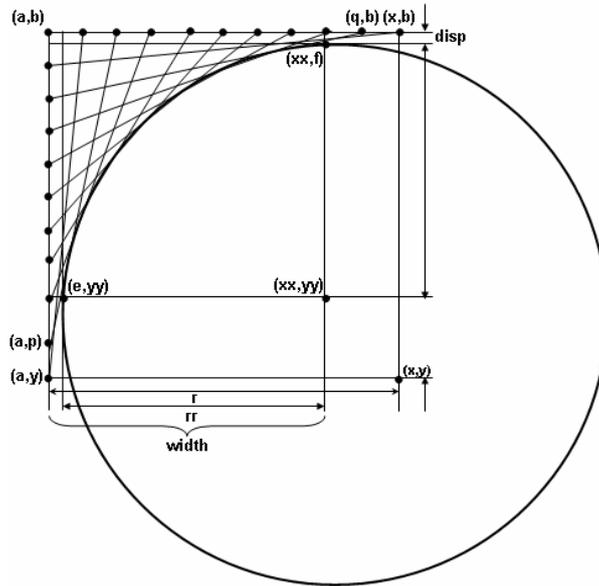

**Fig. 3.** Circle with origin of coordinates at top left corner

The analytical method used to find the transformation is as follows: The origin of coordinates is at the top left corner (refer Fig. 3.)

1. Choose a point $(x, y)$ such that $x = y$ and a distance $r$ and compute the points $(a, y) = (0, y)$, $(x, b) = (x, 0)$ and $(a, b) = (0, 0)$.
2. Divide the horizontal line segment joining $(a, b)$ and $(x, b)$ into $n$ equal parts. Next, divide the vertical line segment joining $(a, b)$ and $(a, y)$ into $n$ equal parts and find the intersection points of straight lines using *Step One* in Section 2. The larger the value of $n$ chosen, the higher is the density of the pixel plotted. The maximum value of $n$ is the number of pixels between the end points of the line segment.
3. With center $(x, y)$ and radius $r$ plot a circle and observe how many of the pixels formed by the intersection of lines fall on the locus of the circle. Compute $(x- i, y- i)$ for $i=1, 2, 3…$ and reduce the radius $r$ until 25% of the 'new' circle pixels coincide with the pixels of intersecting lines. The value of $(x- i, y- i)$ for which the above condition holds true is set to $(xx, yy)$ and the new reduced radius is $rr$. Now that 25% of the circle pixels with center $(xx, yy)$ and radius $rr$ have been found, the remaining 75% of the circle pixels can be found by symmetry about the axes.

With reference to Fig. 3, let $(e, yy)$ be the point of intersection with the pixel on the circle, when a perpendicular is dropped from $(xx, yy)$ onto the line segment joining points $(a, b)$ and $(a, y)$. Similarly, let $(xx, f)$ be the point of intersection with the pixel on the circle, when a perpendicular is dropped from $(xx, yy)$ onto the line segment joining points $(a, b)$ and $(x, b)$. Let the horizontal distance between points $a$ and $e$ be *disp*. By symmetry the vertical distance between points $b$ and $f$ is equal to *disp*.

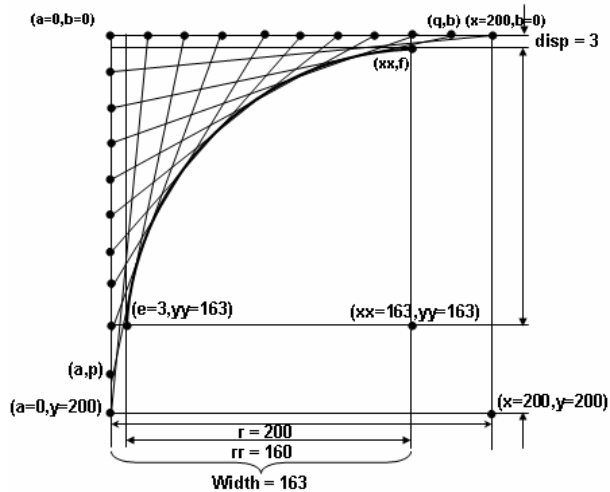

**Fig. 4.** Experimental Setup

Using the analytical method described above, the experiment is setup as follows:
Set $x = 200$, $y = 200$ and $r = 200$. Drop perpendiculars onto the X and Y axes from $(x, y)$ to meet the X axis at $(x, b)$ and Y axis at $(a, y)$ as in Fig. 4. At the co-ordinate $(xx, yy)$ and radius $rr$, it is found that 25% of the locus of the circle coincides with the intersection points of the lines in Fig. 4.

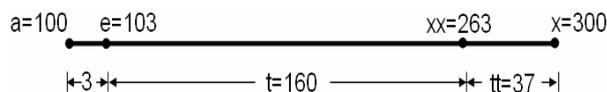

**Fig. 5.** Ratio Derivation

From the experimental setup, the values of $a$, $e$, $t$ and $tt$ are determined and their relationship is as given in Fig. 5. From this setup, it is experimentally determined that $r: rr = 200:160 = 5: 4$. Hence $r = (rr * 5)/4$. Similarly, $t : tt = 160:37$. The distance between $a$ and $e$ gives $disp = 3$. From Fig. 4, it is found that the ratio $(r: dd) = 200:20 = 10:1$, where $dd$ is the vertical distance between $y$ and $p$. Further $x = (rr/160)*3 + (rr/160)*37 + rr$ and $width = (rr/160)*3 + (rr/160) + rr$.

The line intersection pixels between $(e, yy)$ and $(xx, f)$ alone determine the locus of the circle. The points $(a, p)$ and $(q, b)$ are the limiting points for which the line intersections in *Step One* produces circle pixels between $(e, yy)$ and $(xx, f)$ respectively.

Let the distance between $y$ and $p$ is equal to $dd$. Since $(r: dd) = 10: 1$, compute $p = y - (r / 10)$. Similarly, $q = x - (r / 10)$. Now that we know the coordinates of points $(a, p)$, $(a, b)$ and $(q, b)$; one can invoke *Step One* and *Step Two* in section 2 to find 25% of the pixels on the locus of the circle with center $(xx, yy)$ and radius $rr$. The remain-ing 75% of the pixels on the circle is computed using symmetry about the axis. To fill the remaining 75% of the pixels on the circle (with reference to Fig. 3.), for all plotted pixels (*x*, *y*) compute (*x+xoffset*,*y+yoffset*), (2**width-x+xoffset*, *y+yoffset*), (2**width-x+xoffset*, 2**width-y+yoffset*), (*x+xoffset*, 2**width-y+yoffset*). The parameter *xoffset* is the displacement from *x*=0 to *x=a* and *yoffset* is the displacement from *y*=0 to *y=b*, where *xoffset = xx – width* and *yoffset = yy – width*. This is because the simplifications in *Step Two* of section 2 require the point of intersection of the perpendiculars that form the envelope be at (0, 0). Hence, it is shifted to the origin; the necessary simplifications are carried out and then shifted back to its original location. i.e. (a, b).

## 4  The Algorithm

The algorithm takes as input the coordinates of the center of the circle (xx, yy) and radius *rr*. It outputs a circle for the given input radius and center.

The following simplifications are carried out for the parameters in Algorithm 1.
$r = (rr/160)*200 = (rr*5)/4$
$x = (rr/160)*3+(rr/160)*37+rr = (rr*5)/4$
$width = (rr/160)*3 + (rr/160) + rr = (rr*41)/40$
$dist = \lceil (x-(r/10) \rceil +2 = \lceil (10*x - r) / 10) \rceil +2$

Algorithm 1:
Step 1: Input the coordinates of the center of the circle (xx, yy) and radius rr.
Step 2: Compute x←(rr*5)/4, r←(rr*5)/4, width←(rr*41)/40, xoffset ← xx - width, yoffset ← yy - width and dist ← $\lceil (x-(r/10)) \rceil$+2
Step 3: Set $i←\lfloor r/10 \rfloor$+1
   While $i \leq$ dist do
   1. Call Routine 1: find_line_intersection_pixel (0, i-1, x-i+1, 0, 0, i, x-i, 0)
   2. i← i+1

Routine 1: find_line_intersection_pixel ($x_1, y_1, x_2, y_2, x_3, y_3, x_4, y_4$ )
   1. x← ($x_2*(x_2+1)$) / ($y_3+x_4$)
   2. y←($y_3*(x_4-x)$) / $x_4$
   3. Duplicate the pixels in the four quadrants at positions (x+xoffset, y+yoffset), (2*width-x+xoffset, y+yoffset), (2*width-x+xoffset, 2*width-y+yoffset), (x+xoffset, 2*width-y+yoffset)

It is sufficient to plot 12.5% of the pixel points and duplicate the remaining 87.5% pixel points to find the locus of the circle. Algorithm 1 can be easily modified to incorporate these changes and is described in Algorithm 2. The simplifications for parameters *r*, *x* and *width* for Algorithm 2 is same as in Algorithm 1. The parameter *dist* is different from Algorithm 1 and its simplification for Algorithm 2 is as follows
$dist = \lceil ((r/10)+(x-(r/10)))/2 \rceil +2 = \lceil x/2 \rceil +2$

Algorithm 2:
Step 1: Input the coordinates of the center of the circle (xx, yy) and radius *rr*.
Step 2: Compute x←(rr*5)/4, r← (rr*5)/4, width←(rr*41)/40, xoffset←xx-width,
  yoffset←yy-width and dist←⌈x/2⌉+2
Step 3: Set i←⌊r/10⌋+1
  While i ≤ dist do
   1. Call Routine 2: find_line_intersection_pixel
     (0,i-1, x-i+1, 0, 0, i, x-i, 0)
   2. i← i +1
Routine 2: find_line_intersection_pixel ($x_1, y_1, x_2, y_2, x_3, y_3, x_4, y_4$ )
  1. x← ($x_2$*($x_2$+1)) / ($y_3$+$x_4$)
  2. y←($y_3$*($x_4$-x)) / $x_4$
  3. Compute x← x+xoffset and y← y+yoffset.
  4. Plot the pixel (x, y)
  5. Compute x← xx-x, y← yy-y
  6. Duplicate the pixels in the remaining seven quadrants at positions (xx + x, yy + y); (xx - x, yy + y); (xx + x, yy - y); (xx + y, yy + x); (xx - y, yy + x); (xx + y, yy - x) and (xx - y, yy - x)

### 4.1 Error Bound

The rounding off parameters introduces errors in the pixels of the circle plot. The standard equation of a circle with center C (*a*, *b*) and radius *r* is given as $(x - a)^2 + (y - b)^2 = r^2$. The circle plotted using Algorithm 2 with centre CC(xx ,yy) and radius rr; has equation $(x - xx)^2 + (y - yy)^2 = rr^2$. The error is computed as the absolute difference between r and rr. i.e. error = | r –rr |. An idea in [16] on Bresenham's integer only line drawing algorithm, is used for error correction. To correct the error in the direction where the rate of change of y is greater than that of x; set y←y+1 immediately after step 2 of Routine 2 in Algorithm 2.

**Table 1.** Residual Error Table

| Radius (In Pixels) | Average Error | Maximum Error |
|---|---|---|
| 20 | 0.21 | 0.52 |
| 40 | 0.32 | 0.78 |
| 60 | 0.35 | 0.85 |
| 80 | 0.30 | 0.99 |
| 100 | 0.37 | 0.89 |

Table 1 gives the residual errors after error correction for Algorithm 2.

### 4.2 Analysis of the circle algorithm

The two main contributors to the run time of Algorithm 2 are step 1and step 2 of Routine 2. The main operations involved in Routine 2 are in computing

1. $x \leftarrow (x_2*(x_2+1)) / (y_3+x_4)$
2. $y \leftarrow (y_3*(x_4-x)) / x_4$

The multiplications are sped up by employing the modified Booth's algorithm and a Wallace tree (discusses in section 2.1) to give a speed close to addition. The divisions in finding x and y are sped up by a divide and conquer approach on the binary division algorithm and runs in $O(k.\log k)$ time. The division is integer only due to the elimination of floating point operations by suitable use of the floor or ceiling function as discussed in section 2.1. Algorithm 2 plots 12.5% of the points on the locus of the circle and duplicates the remaining 87.5% of the points. In a MIMD environment the generation of each of the 12.5% of the circle points is computed by an independent processor. The proposed algorithm is computationally comparable with the integer only, parallelizable sub arc based algorithm [8] and its improvement [9]. This is clear from the computation of (x, y) pixels in equations 1-4 in [9].

## 5 Conclusion

We propose a parallelizable circle drawing algorithm using integer only arithmetic. Two algorithms, namely Algorithm 1 and its derivative Algorithm 2 with good error bounds (when the radius of the circle is small) are discussed. The operations involved in the individual display of the circle pixels are independent and parallelizable across multiple processors.

## 6 Open Questions

Though the errors from Table 1. are observed to be small, it is seen that the error in pixels for the circle is found to increase when circles with larger radius are required to be drawn. It is believed that these errors are due to the non-coinciding of points of the parabolic arc with the circle as the parabolic arc drawn grows larger. It is not clear if this error can be fixed. If fixable, it can be used as a fast and parallelizable circle drawing algorithm.

## Acknowledgment

The author gratefully acknowledges the suggestions and comments received from Indumathi R, Mike Rosing and anonymous reviewers.

# Source Code

Source code for the Algorithm is available via URL: http://code.google.com/p/cdraw/
Available both in Turbo C on MS-DOS and Allegro gaming library on Linux.

**Screenshots**

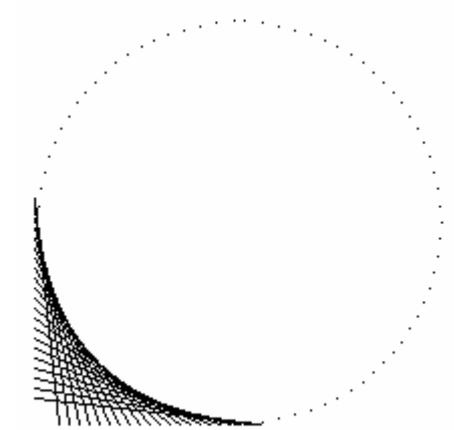

Snap shot 1: The circle drawing algorithm without the invocation of step 2 in section 2

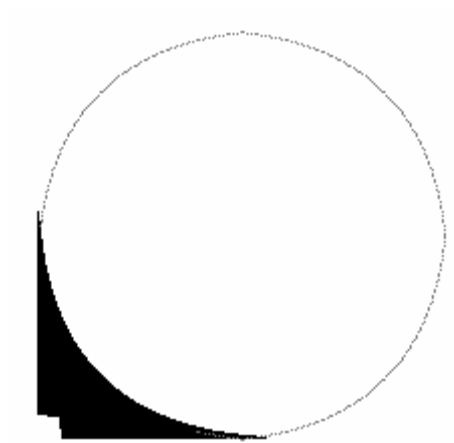

Snap shot 2: The circle drawing algorithm is same as snap shot 1 but with higher pixel density